\documentclass[twocolumn,prd]{revtex4-1}

\usepackage{amsmath}
\usepackage{graphicx}
\usepackage{dcolumn}
\usepackage{bm}
\usepackage{nicefrac}
\usepackage{physdefs}

\newcommand{\HeThree}{\mbox{$^3$He}}
\newcommand{\HeFour}{\mbox{$^4$He}}

\newcommand{\LiSix}{\mbox{$^6$Li}}
\newcommand{\LiSeven}{\mbox{$^7$Li}}

\newcommand{\RA}{\phi}
\newcommand{\Dec}{\theta}

\begin{document}

\title{Is there further evidence for spatial variation of fundamental constants?}

\author{J. C. Berengut}
\author{V. V. Flambaum}
\author{J. A. King}
\author{S. J. Curran}
\author{J. K. Webb}
\affiliation{School of Physics, University of New South Wales, Sydney 2052, Australia}

\date{3 September 2010}

\begin{abstract}

The detection of a spatial variation of the fine-structure constant, $\alpha$, based on study of quasar absorption systems has recently been reported~\cite{webb10arxiv}. The physics that causes this $\alpha$-variation should have other observable manifestations, and this motivates us to look for complementary astrophysical effects. In this paper we propose a method to test whether spatial variation of fundamental constants existed during the epoch of big bang nucleosynthesis. Using existing measurements of primordial deuterium abundance we find very weak indications that such a signature might exist, but the paucity of measurements precludes any firm conclusion. We also examine existing quasar absorption spectra data that are sensitive to variation of the electron-to-proton mass ratio $\mu$ and $x = \alpha^2 \mu g_p$ for spatial variation.

\end{abstract}

\maketitle

\section{Introduction}
\label{sec:intro}

The results of a very large study using data from both the Keck telescope and the Very Large Telescope provide evidence that there is a spatial gradient in the variation of the fine structure constant, $\alpha = e^2/\hbar c$~\cite{webb10arxiv}. In one direction $\alpha$ appears to have been smaller in the past, while in the other direction it appears to have been larger. Briefly, the method compares atomic spectra taken on Earth with spectra seen in absorption systems at high redshift. Any change in $\alpha$ results in well-understood differences between the two spectra. The use of many atoms and ions in a large number of systems ensures good control of systematics~\cite{dzuba99prl,dzuba99pra}.

Astrophysical observations previously suggested that $\alpha$ may have been smaller in the past~\cite{webb99prl,murphy01mnrasA,murphy03mnras,murphy04lnp}. However, these studies all used spectra taken at the Keck telescope in Hawaii, at a latitude of $20^\circ$~N. The recently observed gradient in $\alpha$, which we will refer to as the ``Australian dipole'', has a declination of around $-60^\circ$, which explains why the Keck data, restricted mainly to the northern sky, originally suggested a time-varying $\alpha$ that was smaller in the past. Similar studies, using a much smaller sample from the VLT in Chile (latitude $25^\circ$~S) at first showed a stringent null constraint~\cite{srianand04prl}. More careful analysis of the this sample suggested that the errors should be enlarged by a factor of six~\cite{murphy07prl,murphy08mnras}. The recent study reported in~\cite{webb10arxiv} is the first large-scale analysis of VLT data for $\alpha$-variation.

The detection of this spatial dipole motivates us to reinterpret, in terms of spatial variation, the existing studies of variation of different fundamental constants in other systems. In this paper we test whether further evidence for a dipole can be found in existing measurements of variation in electron-to-proton mass ratio, $\mu = m_e/m_p$, and combinations of fundamental constants such as $x = \alpha^2\mu g_p$, where $g_p$ is the proton $g$-factor. Finally, we consider possible variations in the relative primordial abundances of elements created in the during big bang nucleosynthesis. We develop a model for the detection of a dipole in primordial abundances and its interpretation in terms of spatial variations of fundamental constants during the first few minutes after the big bang.

\section{Model}
\label{sec:model}

We model the spatial variation of a fundamental constant $X$ by
\begin{align}
\label{eq:model}
\frac{\delta X}{X_0} &= d_X\, \Xi(\mathbf{r})\, ,
\end{align}
where $\delta X/X_0 = (X(\vect{r}) - X_0)/X_0$ is the relative variation of $X$ at a particular place $\mathbf{r}$ in the Universe (relative to Earth at $\mathbf{r}=0$). $d_X$ is the strength of the spatial variation, and $\Xi(\mathbf{r})$ describes its geometry. Note that we have assumed that $\delta X/X_0 = 0$ at zero redshift, however this assumption should be tested for each system one measures using the same methods that are used at high redshift (e.g. by using absorbers within our own galaxy).

Since we are fitting a dipole, we use $\Xi(\mathbf{r}) = r \cos\psi$ where $r=ct$ is light-travel distance measured in giga-lightyears and $\psi$ is the angle between the direction of the measurement and the axis of the dipole. $\psi$ requires two parameters to specify the direction of the dipole: right ascension, $\RA$, and declination, $\Dec$, in the equatorial coordinates. Then $\cos \psi$ is given in terms of the direction of the measurement and dipole axis
\begin{align}
\label{eq:Xi}
\Xi(\mathbf{r}) &= r \cos \psi(\RA_d, \Dec_d) \\
\cos \psi &= \cos(\RA - \RA_d)\,\cos\Dec \cos\Dec_d + \sin\Dec \sin\Dec_d\, .\nonumber
\end{align}
We have additionally assumed that the effect of any spatial variation of fundamental constants increases with distance. This is model dependent for large redshifts: we use the standard $\Lambda_\textsl{CDM}$ cosmology parametrized by WMAP5~\cite{hinshaw09apjss} to determine comoving distance $r$. In these coordinates the Australian dipole of $\alpha$-variation found by \cite{webb10arxiv} is
\begin{equation}
\label{eq:ausdipole}
	\frac{\delta\alpha}{\alpha} = (1.10\pm0.25)\E{-6}\, r \cos \psi \, , 
\end{equation}
where $\psi$ is the angle between the direction of the measurement and the axis of the Australian dipole, ($17.4\,(0.6)$~h, $-58\,(6)^\circ$) in equatorial coordinates.

For any given set of data measuring variation of fundamental constants, one should ask whether a dipole model fits the data better than a monopole or a null hypothesis (that is, no variation). However, our goal here is to see if any of the existing data provide further evidence for the spatial variation of fundamental constants seen in~\cite{webb10arxiv}. Therefore it is also reasonable to question whether a dipole model with the axis specified by the Australian dipole provides a better fit to the data. There are some good theoretical justifications for such a procedure. For instance, the constants may vary because they are coupled to a (dimensionless) scalar field $\Phi$ which varies over space-time --- for example, the quintessence field $\Phi/c^2$ or a dimensionless dilaton field. In this case the axis of the dipole is the direction of its gradient $\nabla\Phi$, and a fundamental constant $X$ is coupled to its variation via
\begin{equation}
\frac{\delta X}{X_0} = k_X \delta \Phi\, ,
\end{equation}
where $k_X$ is a dimensionless coupling coefficient. Our dipole model now requires
$\delta \Phi (\mathbf{r}) \sim \Xi(\mathbf{r})$
but all constants will vary in the same direction (i.e. along the Australian dipole).

In this paper we discuss the constants $\alpha$, $\mu$, and the dimensionless mass ratio $X_q = m_q/\Lambda_\textrm{QCD}$, where $m_q$ is the light-current quark mass and $\Lambda_\textrm{QCD}$ is the position of the Landau pole in the logarithm of the running strong coupling constant, $\alpha_s \sim 1/\ln (\Lambda_\textrm{QCD}r/\hbar c)$. In the Standard Model the electron and quark masses are proportional to the vacuum expectation of the Higgs field, while the proton mass $m_p$ is proportional to $\Lambda_\textrm{QCD}$ (if we neglect the $\sim 10\%$ contribution of the quark masses). Relative variation of $X_q$ is then approximately equal to the relative variation of $m_q/m_p$ and $\mu = m_e/m_p$. We can relate the relative variation of different constants by equations like
\begin{equation}
\label{eq:k_mu_alpha}
	k_\mu = R^{\alpha}_{\mu}\, k_\alpha\, ,
\end{equation}
where the $R_X^{X'}$ can be determined from observations and compared with theories of spatial variation.

\section{H$_2$ absorption lines}
\label{sec:mu}

Limits on variation of $\mu$ at high redshift come from comparison of different rotational-electronic excitations in molecular hydrogen (H$_2$). The most recent determinations use the laboratory wavelengths and sensitivity coefficients presented in~\cite{reinhold06prl,ubachs07jms,salumbides08prl} to determine $\delta \mu/\mu$ in four different quasar absorption systems~\cite{king08prl,malec10mnras}.

To determine which model of variation provides the best fit to the $\delta\mu/\mu$ data, we can use $\chi^2/\nu$ as a goodness-of-fit parameter. Here $\nu$ is the number of degrees of freedom; for four data points (quasar absorption systems), $\nu = 4 - p$ where $p$ is the number of parameters in the model. For example, fitting a monopole to the data reduces $\chi^2/\nu$ from 1.21 (null hypothesis) to 1.08 (one parameter), which explains why the data mildly supports a non-zero ``detection'' of $\delta\mu/\mu = 3.4\,(2.7)\E{-6}$.

Alternatively, we can test whether the data supports a dipole model in the direction of the Australian dipole. Using the model~\eref{eq:model} with the dipole direction fixed at (17.4~h, $-58^\circ$) we obtain $d_\mu = 2.6\,(1.3)\E{-6}$/Gyr, with a reduced $\chi^2/\nu = 0.33$. We can compare this to the null hypothesis using an $F$-test, which allows us to find the probability that $\chi^2$ has improved by chance. We define $F$ by
\begin{equation}
\label{eq:F}
	F (p_2 > p_1) = \frac{(\chi^2_1 - \chi^2_2)/(p_2 - p_1)}{\chi^2_2/(n - p_2)}\, ,
\end{equation}
where $p_1$ and $p_2$ are the number of parameters in the respective models and $n$ is the number of data points (in the present case, $p_1 = 0$, $p_2 = 1$, and $n = 4$). Note that we cannot use this test to determine whether the dipole fit is better than the monopole (although $\chi^2/\nu$ suggests that it is), since the models compared using the $F$-test must be ``nested'': it must be possible to generate model 1 from model 2. In any case, since there is no detection the null hypothesis is really the better comparison. We obtain $F = 11.6$ and the probability that this result occurs due to chance is 4\%.

Clearly the Australian dipole provides a good fit to the data, but what does the data alone say? To find the direction preferred by the data, we use the three parameter fit~\eref{eq:model}; the result is $d_\mu = 3.3\,(1.5)\E{-6}$, $\RA_d = 16.7\,(1.5)$~h, $\Dec_d = -62\,(5)^\circ$. This direction is consistent with the Australian dipole. With only one degree of freedom remaining, $\chi^2 = 0.02$. The errors for each model parameter are found by varying the parameter locally near the best-fit value until $\chi^2 = \chi^2_\textrm{min}+1$, while keeping all other parameters fixed at their respective best-fit values. This ignores correlations between the parameters, which must be important; for example if $d_\mu = 0$ (two standard deviations from its best-fit value) then $\RA_d$ and $\Dec_d$ are completely unspecified.

While the preceding paragraphs could be interpreted as evidence for spatial variation of $\mu$, a few things must be noted. Most obviously, we have a very small sample of data, which in principle can be improved although there are significant observational challenges in obtaining data of sufficient quality at high redshift~\cite{curran06mnras,malec10mnras}. Also of importance to this work is the lack of sky-coverage represented in the sample. All four sources are below the equator, and three of them are within $25^\circ$ of each other. Therefore many more data points over a wider sky coverage are needed to obtain a significant detection.

Once a detection is confirmed, we will be able to extract $R^\alpha_\mu$ of \Eref{eq:k_mu_alpha}. For example, with the $d_\mu$ suggested by this work and $d_\alpha$ given by \eref{eq:ausdipole}, $R^\alpha_\mu = 2.6\E{-6}/1.1\E{-6} = 2.4$. However, in \Sec{sec:singles}, we discuss stringent limits from studies of ammonia inversion spectra which, if confirmed, place very strong limits on cosmological variation of $\mu$ and hence $R^\alpha_\mu$. 

\section{21-cm radio vs. UV lines}
\label{sec:x}

The variation of a combination of fundamental constants is probed by comparing neutral hydrogen (\mbox{H\,I}) 21-cm radio and UV metal lines that have been redshifted into the optical band (see~\cite{tzanavaris05prl,tzanavaris07mnras}). The ratio of these transition frequencies is proportional to $x = \alpha^2 \mu g_p$. Only a handful of quasars have been found for which both optical and radio absorption are observed; the nine systems with good data are presented in \cite{tzanavaris07mnras}. These span the redshift range $z=0.23$ to $2.35$.

It was found in \cite{tzanavaris07mnras} that there is far more scatter in the data than would be expected from the statistical errors: the best monopole (single parameter) fit gives $\chi^2/\nu = 8.1$ ($\chi^2=64.6$), where $\sim 1$ would be expected if the errors were Gaussian and the model correct. Therefore \cite{tzanavaris07mnras} concluded that there are large systematic errors, which they trace to the assumption that the 21-cm absorption component is at the same physical location as the UV component. It is supposed that the scatter seen in the data mainly comes from this assumption being incorrect; that is, the 21-cm and UV absorbing gases can be randomly offset in space and have quite different velocities (although see~\cite{curran07mnras,curran10mnras}).

Having enlarged the errors in order to force $\chi^2/\nu = 1$ for the monopole model, it is quite unsurprising that \cite{tzanavaris07mnras} found no evidence for a dipole: the monopole model has been made to fit as well as could possibly be expected. Therefore we have reanalysed the data \emph{without} the additional systematic to test whether any significant dipole exists. Firstly, using the axis of the Australian dipole, we find $\chi^2$ reduces from 110 (zero parameters) to 101 (one parameter). $\chi^2/\nu$ is actually worse than for the monopole fit. Fitting the three parameter dipole \eref{eq:model} provides part of the explanation for why the Australian dipole is such a poor fit: the best fitting dipole is at (7.3~h, $23^\circ$), an angle of $140^\circ$ from the Australian dipole. For this model $\chi^2/\nu = 5.5$, which shows that even with the best-fit dipole systematics still dominate.

We conclude that the detection of a dipole in $x$ of real significance will require much more data than the other systems discussed in this paper, due to the need to account for the large systematic errors caused by the velocity offset between the \mbox{H\,I} and metal absorption lines.

\section{Interpreting single measurements}
\label{sec:singles}

In this section we briefly discuss how individual measurements of combinations of fundamental constants can be compared with the dipole model presented in \Sec{sec:model}. When there are only one or two samples, it is not possible to determine whether a dipole fit to the data is appropriate. However, one can still compare the measurement with the variation that the Australian dipole would lead us to expect.

Two new measurements comparing \mbox{H\,I} 21-cm with neutral carbon (\mbox{C\,I}) absorption lines at $z = 1.36$ and $z=1.56$, along lines of sight to quasars Q2237--011 and Q0458--020, have been reported recently~\cite{kanekar10apjlett}. As discussed in \Sec{sec:x}, these measurements probe the combination of constants $\alpha^2 \mu g_p$. In \Tref{tab:singles} we compare the results of these measurements with the variation expected according to the Australian dipole and the model presented in \Sec{sec:model}. To obtain the model prediction, we calculate the expected variation of $\alpha$ according to \eref{eq:ausdipole} and leave the factors $R^\alpha_{g_p}$ and $R^\alpha_\mu$, defined by equations such as \eref{eq:k_mu_alpha}, for later determination. The type of analysis presented here could eventually yield values for the ratios $R^\alpha_{g_p}$ and $R^\alpha_\mu$, but a larger sample of measurements is required.

\begin{table*}[t]
\caption{\label{tab:singles} Comparison of expected variation, given by \Eref{eq:ausdipole}, and measured variation of fundamental constants in different systems. Each measurement corresponds to a single absorption system. $R^\alpha_{g_p}$ and $R^\alpha_\mu$ are defined by equations like \eref{eq:k_mu_alpha}.}
\begin{ruledtabular}
\begin{tabular}{llccc}
System & Constant & Expected variation & Measurement & Ref. \\
       &          &$(\times 10^{-6})$ & $(\times 10^{-6})$ \\
\hline
\mbox{H\,I} 21-cm + \mbox{C\,I} & $\alpha^2 \mu g_p$
    & $1.12\times(2 + R^\alpha_\mu + R^\alpha_{g_p})$
    & $6.64 \pm 0.84_\textrm{stat} \pm 6.7_\textrm{sys}$ & \cite{kanekar10apjlett} \\
 &
    & $-5.20\times(2 + R^\alpha_\mu + R^\alpha_{g_p})$
    & $7.0 \pm 1.8_\textrm{stat} \pm 6.7_\textrm{sys}$ & \cite{kanekar10apjlett} \\
NH$_3$ inversion & $\mu$
    & $-5.47\ R^\alpha_\mu$
    & $< 1.8\E{-6}\ (2\sigma)$ & \cite{murphy08sci} \\
 &
    & $1.34\ R^\alpha_\mu$
    & $< 1.4\E{-6}\ (3\sigma)$ & \cite{henkel09aap} \\
\mbox{H\,I} 21-cm + OH 18-cm & $(\alpha^2/\mu)^{1.57} g_p$
    & $-1.04\times(3.14 - 1.57 R^\alpha_\mu + R^\alpha_{g_p})$
    & $4.4 \pm 3.6_\textrm{stat} \pm 10_\textrm{sys}$ & \cite{kanekar05prl} \\
OH 18-cm & $(\alpha^2/\mu)^{1.85} g_p$
    & $0.50\times(3.70 - 1.85 R^\alpha_\mu + R^\alpha_{g_p})$
    & $-11.8 \pm 4.6$ & \cite{kanekar10apjlett0} \\
\end{tabular}
\end{ruledtabular}
\end{table*}

Measurements of $\mu$-variation have been made by comparing inversion lines of ammonia, NH$_3$, with rotational lines of CO, HCO$^+$, and HCN molecules. The inversion spectrum of ammonia has an enhanced sensitivity to variation of $\mu$ because it depends on the exponentially small probability of tunneling of the three hydrogen atoms through the potential barrier~\cite{flambaum07prl}. In \Tref{tab:singles} we present two measurements of $\mu$ variation based on this method.

Two other measurements are shown in \Tref{tab:singles}. The combination $(\alpha^2/\mu)^{1.57} g_p$ can be extracted by comparison of OH 18-cm and \mbox{H\,I} 21-cm lines; this has been performed for the gravitational lens toward PMN J0134--0931 at $z = 0.765$~\cite{kanekar05prl}. A measurement of $(\alpha^2/\mu)^{1.85} g_p$, derived from comparison of conjugate-satellite OH 18-cm lines, has yielded a $2.6\sigma$ non-zero detection towards PKS 1413+135~\cite{kanekar10apjlett0}. We note that this absorber lies at $\sim 81^\circ$ to the Australian dipole and is at low redshift ($r \sim 2.9$~Gyr), therefore minimal variation in fundamental constants is expected.

\section{Big bang nucleosynthesis}
\label{sec:bbn}

In principle one can search for spatial variation in the primordial abundance of any element produced during BBN (the most common are D, $^3$H, \HeThree, \HeFour, \LiSeven). The best limits on primordial deuterium abundance are derived from isotope-shifted Ly-$\alpha$ spectra in quasar absorption systems, where the absorbers are at $z\sim 2.5$. By contrast \HeFour\ is observed in ionized (\mbox{H\,II}) regions of low-metallicity dwarf galaxies at $z \lesssim 0.01$, while the primordial \LiSeven\ abundance is determined from metal-poor Population II stars in our galaxy. Therefore, in practice only the deuterium abundance is currently measured at large enough redshifts to expect to see any effect of spatial variation of fundamental constants.

In the case of primordial abundances we need to extract an average value from the data (monopole term) as well as any potential dipole. Therefore we use the model
\begin{align}
\label{eq:D_model}
\log_{10}(a/\textrm{H}) &= m_a + d_a\, \Xi(\mathbf{r}) \\
    &= m_a + d_a\, r \cos\psi(\RA_d, \Dec_d)\, , \nonumber
\end{align}
where $\log_{10}(a/\textrm{H})$ is the primordial abundance of element $a$ relative to hydrogen abundance at a particular place $\mathbf{r}$, $m_a$ is the average primordial abundance, $d_a$ is the strength of the spatial variation, and $\Xi(\mathbf{r})$ describes the geometry of the spatial variation \eref{eq:Xi}.

In the future one might hope to measure primordial abundances of the other elements at high redshift. The production of each of these elements has a different sensitivity to fundamental constants, therefore a complete set of data can, in principle, simultaneously measure several fundamental constants at the time of nucleosynthesis. The dependence of primordial abundances on fundamental constants such as $\alpha$ and $m_q/\Lambda_\textrm{QCD}$ (the ratio of light-quark mass to the pole in the running strong coupling constant) is the subject of current research~\cite{dmitriev04prd,flambaum07prc,dent07prd,berengut10plb}. An observed spatial variation in primordial abundance (non-zero $d_a$) can be related to the variation of a fundamental constant $X$ at that position in space at the time of big bang nucleosynthesis using the relationships
\begin{equation}
\label{eq:interpret}
d_a = \frac{\partial \log_{10}(a/\textrm{H})}{\partial \Xi}
    = \frac{\partial \log_{10}(a/\textrm{H})}{\partial \ln X}\,\frac{\partial\ln X}{\partial\Xi}\, ,
\end{equation}
where $\partial\log_{10}(a/\textrm{H})/\partial\ln X$ is determined from theory.

We have performed a preliminary search for such a dipole using the seven best determinations of primordial deuterium abundance presented in \cite{pettini08mnras}. It is known that the standard (monopole) model gives a minimum $\chi^2$ of $19.13$ rather than the $\sim 6$ that would be expected if the errors were purely statistical, therefore there is at least a possibility that the dipole model of spatial variation could produce a better fit to the data. As in the cases presented in earlier sections, we can investigate this possibility using $\chi^2/\nu$ as a measure of the fitness of our model. The results are shown in \Tref{tab:D_results}.
We find that the dipole model with direction fixed by the Australian dipole~\cite{webb10arxiv} is not significantly preferred over the simple monopole model, with $\chi^2/\nu$ increasing very slightly. On the other hand, if we let the data \emph{choose} the dipole direction, in effect solving for $m$, $d$, $\RA_d$ and $\Dec_d$ simultaneously, we only have three remaining degrees of freedom, and $\chi^2/\nu$ increases substantially. Interestingly, the dipole is found to point in the same direction (within errors) as the Australian dipole (see \Tref{tab:D_results}).

\begin{table*}
\caption{\label{tab:D_results} Calculated best fits to the primordial deuterium abundance data using various models. The model parameters, including $m_D$, $d_D$ and the direction of the dipole (right ascension and declination) from \eref{eq:model}, as well as minimum $\chi^2$ and $\chi^2$ per degree of freedom ($\nu$) are shown.
The monopole model is the standard interpretation ($m_D$ only). The dipole model was tested using fixed and varying dipole directions. The fixed dipole direction corresponds to the best fit of \cite{webb10arxiv}: \mbox{(17.4~h, $-58^\circ$)} in equatorial coordinates. Errors in each parameter are found ignoring correlations; see \Sec{sec:mu} text for details. }
\begin{ruledtabular}
\begin{tabular}{lcccccc}
 & $m_D$ & $d_D$ & R.A.    & Decl. & $\chi^2$ & $\chi^2/\nu$ \\
 &       &       & (hours) & (deg) \\
\hline
Monopole & $-4.55\,(2)$ & & & & 19.13 & 3.19 \\
Dipole (fixed direction) & $-4.55\,(2)$ & $0.0045\,(35)$ & & & 17.50 & 3.50 \\
Dipole   & $-4.56\,(2)$ & $0.0054\,(29)$ & $15.5\,(1.6)$~h & $-14\,(51)^\circ$ & 15.73 & 5.24 \\
\end{tabular}
\end{ruledtabular}
\end{table*}

\begin{figure}
\caption{\label{fig:D_data} Deuterium abundance vs. $r\cos\psi$ where $\psi$ is the angle between the measurement direction and the dipole of Ref.~\cite{webb10arxiv}. Dashed line: monopole fit; solid line: dipole fit (fixed direction). Fit parameters are given in \Tref{tab:D_results}.}
\includegraphics[width=0.92\columnwidth]{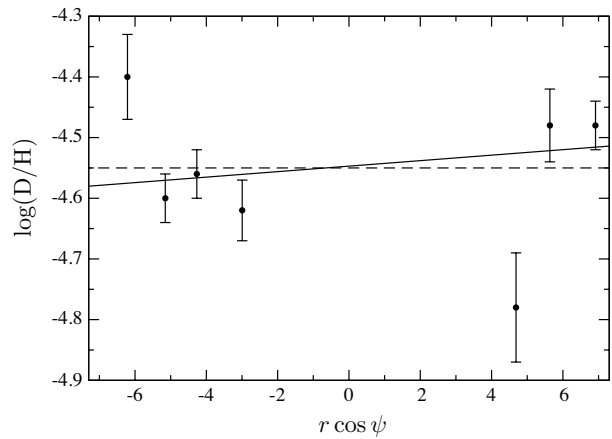}
\end{figure}

In \Fig{fig:D_data} we show the measured abundances against $r\cos\psi$. The dipole fit (solid line) fits better than the monopole fit (dashed line). We clearly need more sky coverage: all of the data points lie at angles $55^\circ < \psi < 125^\circ$ from the dipole axis. The distances are similar for all the data points ($r \sim 11$~Gyr), therefore our results do not vary significantly if we remove the distance dependence and let $\Xi(\mathbf{r}) = \cos\psi$. We note in passing that if the two outlying points in \Fig{fig:D_data} are removed and only five remain, $\chi^2/\nu = 2.00$ for the monopole fit and $0.44$ for the dipole fit: $m = -4.55\,(2)$ and $d = 0.010\,(4)$.

A more reasonable approach to the $\chi^2$ discrepancy is to increase the error bars to account for an unknown systematic. We have done this by adding a constant error $\sigma_\textrm{sys}$ to all the statistical errors in quadrature: \mbox{$\sigma_\textrm{tot}^2 = \sigma_\textrm{stat}^2 + \sigma_\textrm{sys}^2$}. Increasing the errors such that the dipole fit has $\chi^2 = 5$ requires an additional systematic of $\sigma_\textrm{sys} = 0.11$ to be added to all the data points. An increase in the uncertainty of this magnitude destroys any justification for fitting a dipole to the data.

Finally, we interpret our ``detection'' of a dipole in the primordial deuterium abundance, aligned with the Australian dipole, in terms of variations in fundamental constants. The parameter $d_D = 0.0045\,(35)$ and we may interpret this in terms of variation of $\alpha$. Using the result $\partial\ln(\textrm{D}/\textrm{H})/\partial\ln\alpha = 3.6$~\cite{dent07prd} in \Eref{eq:interpret} we obtain
\[
\frac{\delta\alpha}{\alpha} = \frac{0.0045}{3.6/\ln 10}\ \Xi(\textbf{r}) = 0.003\ \Xi(\textbf{r})\, .
\]
In other words, a system 1~Gyr in the direction of the dipole axis had a fine-structure constant different from the present-day laboratory value by $\delta\alpha/\alpha = 0.003$ at the time of big bang nucleosynthesis.

This effect is much larger than that seen directly in quasar absorption spectra: $\delta\alpha/\alpha \sim 10^{-6}\ \Xi$ in \cite{webb10arxiv}. Yet even if this signal is confirmed, there is no inconsistency here. The direct measurements of \cite{webb10arxiv} probe physics occurring a few billion years ago, while the primordial abundance measurements probe the first few minutes after the big bang.

We chose the fine-structure constant since spatial variation of $\alpha$ has been reported. However big bang nucleosynthesis is more sensitive to $X_q = m_q/\Lambda_\textrm{QCD}$, the dimensionless ratio of light-quark mass to the pole in the running coupling constant, and it is generally more reasonable to interpret the variation in terms of smaller shifts in these constants. For example using the result $\partial\ln(\textrm{D}/\textrm{H})/\partial\ln X_q = 7.7$~\cite{flambaum07prc} we obtain
\[
\frac{\delta X_q}{X_q} = 0.0013\,(10)\ \Xi(\textbf{r})\, .
\]

Clearly the detection and confirmation of spatial variation in fundamental constants at the time of big bang nucleosynthesis requires a large improvement in the number of deuterium measurements and their coverage of the sky. However, it may also be possible to detect other elements produced in BBN such as \HeThree, \HeFour, \LiSix, and \LiSeven\ in high-redshift quasar absorption spectra and determine their primordial abundances. If the response of BBN to variation of constants is known, one can readily predict the spatial variation of their abundances given a model of fundamental constant variation. We reproduce some of the known response functions from the theory~\cite{dent07prd,flambaum07prc} in \Tref{tab:BBN_sensitivity}.

\begin{table}
\caption{\label{tab:BBN_sensitivity} The sensitivity of relative variation in primordial abundances to relative variation in fundamental constants $\partial \ln Y_a/\partial \ln X$. Here $Y_a = ($D/H, \HeThree/H, $Y_p$, \LiSix/H, \LiSeven/H$)$ are number ratios of primordial isotope abundances to hydrogen, except $Y_p$ which is the mass ratio \HeFour/H.}
\begin{ruledtabular}
\begin{tabular}{lrrrrrc}
 $X$  & D/H & \HeThree/H & $Y_p$ & \LiSix/H & \LiSeven/H & \\
\hline
$\alpha$ & 3.6 & 0.95 & 1.9 & 6.6 & $-11$ & \cite{dent07prd} \\
$X_q$ & 7.7 & --- & $-0.95$ & --- & $-50$ & \cite{flambaum07prc} \\
$\eta$ & $-1.6$ & $-0.57$ & 0.04 & $-1.5$ & 2.1 & \cite{dent07prd} \\
$G$ & 0.94 & 0.33 & 0.36 & 1.4 & $-0.72$ & \cite{dent07prd} \\
\end{tabular}
\end{ruledtabular}
\end{table}

We can see, for example, that a variation of 1\% in $\alpha$ would amount to a $1.9\%$ change in $Y_p$ (the mass ratio \HeFour/H -- for all other elements the number ratio is presented) and a $-11\%$ change in \LiSeven\ abundance. This provides an independent means of verifying a particular change in fundamental constants.
The primordial \LiSeven\ abundance evidently provides a particularly sensitive probe of fundamental constants. In \Tref{tab:BBN_sensitivity} we also present the sensitivity of primordial abundances to the baryon-to-photon ratio $\eta$ (relative to the WMAP value $\eta = 6.1\E{-10}$), which prior to WMAP was constrained chiefly by BBN data.

\section{Conclusion}

The existing H$_2$ data from quasar absorption spectra shows hints that there may be a dipole in $\mu$-variation with an axis corresponding to that of the gradient in $\alpha$ found in \cite{webb10arxiv}. On the other hand $x=\alpha^2\mu g_p$ variation, taken from \mbox{H\,I} 21-cm data, has a best-fit dipole whose axis does not correspond to that of the Australian dipole, although in this case systematics heavily dominate.
We have demonstrated that it is possible to infer spatial variation of fundamental constants during big bang nucleosynthesis from high-redshift measurements of primordial abundances. Although the existing deuterium data does not necessarily support the dipole interpretation, with the significance of the dipole model being similar to that of the monopole model, it is interesting that the preferred axis is consistent with the direction of the Australian dipole. There is a strong impetus now to perform measurements of relative primordial abundance at high redshifts of as many elements as possible in as many different spatial directions as possible.

Finally, we note that it may be possible to observe a spatial dipole in other cosmological systems~\cite{berengut10arxiv0}. For example, $\alpha$-variation may be seen in the CMB anisotropy if a high-enough sensitivity can be reached. Although the results of \cite{webb10arxiv} (interpreted as strictly spatial variation) suggest that accuracy at the level $10^{-6}$ will be required, if the hints from BBN turn out to be real then there is an additional redshift (time)-dependence that could increase the variation at the time of the CMB substantially. Another possibility is that if the observed $\alpha$-variation is related to the cosmological constant, and hence the accelerated expansion of the Universe, it may be possible to see a dipole in the redshift-luminosity relationships of SnIa supernovae data (see, e.g.~\cite{cooke10mnras}).

We thank M. T. Murphy for useful discussions. This work is partly supported by the Australian Research Council.

\bibliography{references}

\end{document}